\documentclass[useAMS,usenatbib]{mn2e}
\usepackage{graphics,graphicx,epsfig,psfig}
\usepackage[normalem]{ulem}
\usepackage[dvipsnames]{color}
\usepackage{soul,xcolor}
\usepackage{amsmath, amssymb}
\usepackage[]{inputenc,amssymb}
\usepackage{booktabs,caption}
\usepackage{amsmath}
\usepackage{multirow}
\usepackage{soul}

\setstcolor{red}

\def \be{\begin{equation}}
\def \ee{\end{equation}}
\def \ba{\begin{eqnarray}}
\def \ea{\end{eqnarray}}
\def \etal{{et al.}}

\definecolor{webgreen}{rgb}{0,.5,0}
\definecolor{webbrown}{rgb}{.6,0,0}

\usepackage[%
    colorlinks = true,%
    linkcolor = blue,%
    urlcolor  = blue,%
    citecolor = webgreen,%
    anchorcolor = blue]{hyperref}

\newcommand{\ufhref}[3][blue]{\href{#2}{\color{#1}{#3}}}%

\title[Radio background and IGM heating from Pop III SN]{Radio background and IGM heating due to Pop III supernovae explosions}
\author[Jana, Nath, Biermann]
{Ranita Jana$^1$\thanks{E-mail: ranita@rri.res.in}, Biman B. Nath$^1$, Peter L. Biermann$^{2,3}$\\
$^1$ Raman Research Institute, Sadashiva Nagar, Bangalore 560080, India\\
$^2$ Max Planck Institut f\"ur Radioastronomie, Auf dem H\"ugel 69, Bonn, Germany\\
$^3$ Department of Physics, Karlsruhe Institut f\"ur Technologie, Karlsruhe, Germany\\
}
\begin{document}
\maketitle
\label{firstpage}
\begin{abstract}
We consider the synchrotron emission from high energy electrons accelerated in supernova explosions of massive Population III stars in high redshift minihaloes of mass $10^{5\hbox {--}7} \rm M_\odot$. We show the resulting intensity of radio background from this process can be substantial, which could potentially explain the recently reported {\it EDGES} result, if not for the associated heating of the IGM by CR protons which are also produced at the same time. The trade-off between the radio background and heating is such that the 21 cm brightness temperature cannot be larger than $\vert \Delta T_{21}\vert \sim 0.25$ K. The radio background and heating are both produced by energetic particles, although one by energetic electrons and the other by energetic protons. The two competing processes, production of radio background and heating of IGM by Pop III supernovae, determine the depth of the trough in the 21 cm brightness temperature which can be observed in future experiments and used as a test of this scenario.
\end{abstract}
\section{Introduction}
\label{sec:intro}
The epoch in the history of the Universe when the first luminous objects formed has long been a topic of interest. It has been estimated that the Population III (Pop III) stars formed inside dark matter haloes (the so-called minihaloes) with virial temperature $\sim 1000$ K and mass $\sim 10^6$ M$_\odot$ at $z\sim 20\, - \,30$ \citep{haiman1996, tegmark1997, bromm2009}. They fundamentally transformed the Universe not only by producing the first sources of light but also in other aspects. 
The ionizing radiation from these luminous objects (perhaps aided by those from early quasars usually interpreted as activity of super-massive black holes \citep{puchwein2018}; for the activity of early stellar mass black holes, see e.g., \citep{Mirabel2011}) ultimately led to the reionization of the Universe. This ionizing radiation
could have come from either the Pop III stars themselves \citep{venkatesan2003} or from the Pop III supernovae explosions \citep{johnson2011}.
The supernovae explosions likely accelerated cosmic rays (CR), which would have heated the intergalactic medium (IGM) \citep{sazonov2015}. These CRs could have also produced a neutrino background, as has been estimated by \citet{berezinsky2012}.

A powerful probe of these early epoch is the redshifted 21 cm radiation from neutral hydrogen atoms \citep{Barkana2001}. The properties of this radio signal strongly depend on the deviation of the HI spin temperature from the cosmic microwave background (CMB) temperature ($\Delta T_{\rm 21} \propto  (T_S -T_{CMB})$). The kinetic temperature of the inter-galactic medium was coupled to the HI spin temperature by $z\sim 15$ due to the resonant scattering of Ly$\alpha$ photons. This gives rise to a prominent absorption feature in the global 21 cm signal. None of the existing reionization models can explain the observed absorption trough of the first detected redshifted global 21 cm spectrum at $z\sim 17$, \citep{bowman2018}  the depth of which is almost twice than expected. The shape of the absorption trough with its sharp edges is also rather difficult to understand in the context of standard models. It should be noted that the detection of first redshifted  global 21 cm signal is yet to be confirmed. \cite{hills2018} expressed concerns about the cosmological origin of this signal.

There have been suggestions (e.g., \citep{barkana2018, barkanaetal2018, berlin2018}; see also \cite{biermann2006}) that there might be baryonic-dark matter interaction which would have caused excess cooling of the cosmic gas,
leading to a deep absorption trough. Another possibility is that of a global radio background which would also explain the observations \citep{feng2018}. Indeed, such a radio background at $z\sim 20$ was predicted by \citet{biermann2014}, arising from explosions concomitant with the formation of first supermassive black holes, motivated by the excess radio background observations at the present epoch \citep{fixsen2011}, which has been independently confirmed recently by \citet{Dowell2018}. Recently, \cite{ewallwice2018} considered 
the radio background from growth of seed black holes at high redshift, and determined the required black hole seed function. However, the physics of the formation
of these seed black holes remain highly uncertain.

In this paper, instead of invoking early black holes, we consider another, and perhaps more abundant, source of radio background, namely, the effect of supernovae (SNe) explosions of Pop III stars in dark matter minihaloes  which were abundant at these redshifts. We have estimated the brightness temperature of the radio background generated by high energy CR electrons interacting with magnetic fields in the shocked inter-galactic medium, and shown how this radio background could have been important vis-a-vis the 21 cm absorption experiment.

\section{Pop III supernovae and cosmic rays}
It is believed that Pop III stars appeared at $z \sim 20\, - \,30$ in dark matter minihaloes of mass $10^{5\hbox {--}7} \rm M_\odot$. Although there are significant uncertainties in the initial mass function of these stars, it is thought that the mass function was dominated by massive stars in the range of $\sim 10\, - \,10^3 \rm M_\odot$. They lived for several million years (e.g. \cite{schaerer2002}) and a significant fraction (or most) of them likely exploded as powerful pair-instability SNe, with energies much larger than that of present day SNe, of the order of $\sim 10^{53}$ erg (\cite{sazonov2015},\cite{hirano2014},\cite{heger2010}).
It has also been shown that the UV radiation from Pop III stars forms HII regions, and the supersonic shock waves associated with R-type ionization front sweeps away most of the gas in the minihaloes (e.g.\cite{yoshida2007}). This decrease in gas density reduces the radiation loss of subsequent SN remnants (SNR), which are able to travel beyond the virial radius of the minihaloes. They would suffer radiation loss when they encounter the shell of gas previously blown away by the ionization front. However, it has been found
that these primordial SNRs would travel to a few times the virial radius before mixing with the IGM, and thereby destroying the minihaloes of mass $\le 10^7$ M$_\odot$, the
range we are interested in here
 \citep{kitayama2005, vasiliev2008}. We note that these explosions would leave no stellar remnants and therefore the scenario under consideration here does not involve black holes of any mass range.
 
 Consider the SNRs from Pop III stars around minihaloes at $z\approx17$. The minimum mass of halos that can efficiently cool with molecular hydrogen is $\approx 5 \times 10^5$ M$_\odot$ (for $h=0.7$) \citep{tegmark1997, yoshida2003}.
 If we consider minihaloes in the mass range $5\times 10^5\, -\,10^7$ M$_\odot$, then at $z= 17$, their comoving number density is $n_h \approx 338 (h/0.6774)^3$ Mpc$^{-3}$, using the {\it CAMB} transfer function calculator and the fitting function of \citet{reed2007}. This has been calculated using the HMF calculator given by \citet{murray} and the cosmological parameters used are determined by \citet{planck}. Suppose that each minihalo gives rise to $f_{\rm SN}\sim 1$ SN with energy $E_{\rm SN}\sim 10^{53} E_{53}$ erg. The baryonic mass density of the IGM is
$\rho_{\rm IGM}(z)\approx 2.4 \times 10^{-27} \Big(\frac{f_b}{0.157}\Big) \Big(\frac{\Omega_{m0}}{0.3089}\Big) \Big(\frac{1+z}{18}\Big)^3 $ g cm$^{-3}$, for a cosmic baryon fraction of $\sim 0.157$ and the matter density fraction at present epoch $\Omega_{m0}\sim 0.3089$.
 There are three distinct stages of SNR evolution. At first the ejecta freely expands until the swept out mass is comparable to the ejecta mass. It is followed by an adiabatic expansion of the shocked gas which is known as Sedov-Taylor(S-T) phase. The (physical) radius in the S-T phase is,
 \be
 R_{ST}\sim 1.2 \, {\rm kpc} \, E_{53}^{1/5} \, t_{14}^{2/5} \, \rho_{\rm IGM}(z)^{-1/5}\,,
 \ee
 where $t_{14}$ is time after the explosion in the unit of $10^{14}$ s $\approx 3$ Myr. In comparison, the virial radii of minihaloes of the considered masses are in the range $\sim 143\,- \,389$ pc. These SNRs lose energy
 through inverse Compton scattering off CMB photons, with a cooling time $t_{\rm IC}\sim 7\times 10^{14}$ s at $z\approx 17$ for Lorentz factor $\gamma_e=1$. 
 However, the radiative phase begins at $t\sim 10^{14}$ s, somewhat earlier, when the cooling of the shocked gas reaches a maximum rate \citep{sutherland1993}. This happens when the post shock temperature of the ionised shell is $\approx 1 \hbox{--} 2 \times 10^5$ K,  corresponding to the shell speed of order $v\sim 90 \hbox{--} 120\, E_{53}^{1/5} \rho_{\rm IGM}(z)^{-1/5} t_{14}^{-3/5}$ km s$^{-1}$. This is therefore the epoch of beginning of the radiative phase.
 Thereafter the SNRs rapidly decelerate and disperse into the IGM. 
  The spacing between minihaloes, from the above mentioned number density, is estimated as $\sim 4.9$ kpc. In other words, the SNRs lose steam by the time they reach a distance $\sim 25\%$ of the inter-halo distance. The volume filling fraction is $\approx 0.02$.
 
 Like Galactic SNRs, these primordial SNRs in all likelihood accelerate CRs by diffusive shock acceleration (DSA). The prerequisite for this method to work are magnetic irregularities and the formation of shocks. The magnetic field in the IGM at these epochs is likely to be scaled up from the present day value by a factor of $(1+z)^2$.
Observations suggest that  $B\le 10^{-9}$ G at present epoch (although this is valid for a coherence length of $\sim 1$ Mpc; \citep{kronberg1994, subramanian2016}). Therefore one expects
an intergalactic magnetic field of order $B_{\rm IGM}\sim 0.32\, \mu$G at $z\approx 17$. One possible origin of this magnetic field could be stellar winds and SNe of Pop III stars themselves \citep{bisnovatyikogan1973}. These seed magnetic fields can be amplified by small-scale dynamo in the cores of minihaloes \citep{sur2010}.

Recall that the surroundings of minihaloes are expected to be ionised by
the UV radiation of Pop III stars. Simulations of the ion Weibel instability even in the absence of pre-existing magnetic fields have shown that a shock front can
form in this case \citep{Weibel1959, spitkovsky2008}. 
Although these simulations have been carried out for relativistic plasmas, it is likely that the result will hold for non-relativistic motions of Pop III SNRs since the general physical principles that lead to the formation of shock fronts remain unchanged \citep{berezinsky2012}.

The next requirement for DSA is the existence of magnetic irregularities upstream and downstream of the shock. It is thought that cosmic ray streaming instability can cause 
magnetic turbulence upstream. It is a non-linear process, and accelerated particles excite the instability while DSA occurs in the presence of turbulent magnetic field
\citep{Lucek2000, Bell2001}.
\citet{berezinsky2012} estimated that magnetic field in the vicinity of Pop III SNRs would reach  up to $\delta B \sim 4.7 \, V_9^{1/2}\, \, \mu$G, using an ambient 
magnetic field of $0.32\mu$G, IGM particle density $n_{\rm IGM}\sim 2.4 \times 10^{-3}$ cm$^{-3}$ and an efficiency of CR acceleration (see below) of $0.15$ in their equation 3 
when the non-linearity proceeds in a resonant way. Here $V_9$ is shock speed in the units of $10^4$ km s$^{-1}$. This implies an amplification $\delta B/B \sim 14.7 \, V_9^{1/2}$. 
In the non-resonant case, their equation 4 leads to an amplification factor of $\delta B/B \sim 35 \, V_9^{3/2}$. 
For downstream magnetic fields, observations  indicate that the magnetic field in young SNRs (in our Galaxy \citep{bell2013}, or in other galaxies, e.g, in M82 
\citep{biermann1985}) can be as large as $\sim 10\, - \,100$ times the ISM value, or even more (see Table 1 of some individual radio SNRs in \cite{biermann2018}).  
We will assume a fiducial value of $\delta B/B \sim 100$ and characterise the factor
by which magnetic field is increased from the IGM value by $\zeta_B\sim 100 \, \zeta_{B, 100}$.

The efficiency with which the shock kinetic energy is converted into CR energy is another uncertainty. Simulations of DSA in non-relativistic shocks show that a fraction
$\eta_{\rm CR} \sim 0.1\, - \,0.2$ of the shock kinetic energy is spent in accelerating CRs \citep{caprioli2014}. As a fiducial value we will assume $\eta_{\rm CR}\sim 0.15$. 
Since most of the CR energy is in protons, we need another parameter to characterise the spectrum of CR electrons, namely, the ratio of CR electrons to protons energy. 
We assume that this ratio is given by $(m_e/m_p)^{(3-p)/2}$, where $p$ is the power-law index of the energy spectrum of CRs (see equation 10 in \cite{Persic2014}; also
\cite{merten2017}). We will assume $p\approx 2.2$, \citep{berezinsky2012} for which
this ratio is $\eta_e\approx 0.05$. 
Together with the efficiency parameter $\eta_{\rm CR}$, the number density of minihaloes $n_h$ and the energy of Pop III SNe, $E_{\rm SN}$, the energy 
density of CR electrons (in physical units) is therefore,
\ba
\epsilon_{\rm CR, e}&=&\eta_e \eta_{\rm CR} f_{\rm SN} E_{\rm SN} n_h\nonumber\\
&&=0.05 \times 0.15 \times 1 \times 10^{53} \rm erg \times \frac{338 \, (1+17)^3}{Mpc^3}\, \nonumber\\ &&\approx 5 \times 10^{-17} \rm erg \, cm^{-3}  \,.
\ea
Now if $n(\gamma_e) d\gamma_e$ denotes the number density of electrons with Lorentz factor $\gamma_e$ in the range $\gamma_e$ to $\gamma_e + d\gamma_e$ then this allows us to write the CR electron energy spectrum as $n(\gamma_e) d\gamma_e=A \gamma_e ^{-p} d\gamma_e$, with the normalization,
\be
A={(p-2) \epsilon_{\rm CR,e} \over m_e c^2}=6.1 \times 10^{-11} \, (p-2)  \, E_{53} \, {\rm cm}^{-3} \,.
\ee
Here the units are physical, and we have used the fiducial values for $\eta_e, \eta_{\rm CR}, f_{\rm SN}$ mentioned above.

The synchrotron emissivity of these CR electrons in magnetic field $B$ averaged over all electron pitch angles is given by \citet{rybicki2004},
(with $\Gamma$ denoting gamma functions)
\ba
j_\nu&=&
\frac{\sqrt 3 e^3 A B} {8 \sqrt{\pi} m_e c^2 (p+1)} \Gamma \Big(\frac{p}{4} + \frac{19}{12}\Big) \Gamma \Big(\frac{p}{4}-\frac{1}{12}\Big) \,
\frac{\Gamma (\frac{p+5}{4})}{\Gamma (\frac{p+7}{4})}\,
\nonumber\\ 
&& \times \Big(\frac{2 \pi m_e c \nu}{3 e B }\Big)^{-(p-1)/2}
\nonumber\\
&\approx&  3.1 \times 10^{-43} \, {\rm erg} \, {\rm s}^{-1} \, {\rm Hz}^{-1} \, {\rm cm}^{-3} \, {\rm sr}^{-1} \, \zeta_{B,100}^{1.6} \, E_{53} \,.
\ea

The specific intensity is given by $I_\nu=j_\nu \times (c/H(z))$. We have $c/H(z)\approx 104$ Mpc at $z=17$. Therefore
we have,
\be
I_\nu\approx  10^{-16} \, \zeta_{B,100}^{1.6} \, E_{53}\, \frac{\rm erg \hbox{  } s^{-1}}{\rm cm^2 \hbox{ }sr \hbox{ } Hz}
\ee
The corresponding brightness temperature is
\be
T_B(z=17)= \frac{I_\nu c^2}{2 k \nu^2} =160 \, \zeta_{B,100}^{1.6} E_{53} \hbox{ }\rm K \,.
\ee
This increases the depth of the absorption trough by a substantial factor $={(T_B+T_{\rm CMB})\over T_{\rm CMB}}$ from the standard value. Therefore,
given the 
uncertainties and the dependence on $B$, a radio background that can explain the observed absorption trough is tenable.

\begin{figure}
\centering
\includegraphics[height= 2in]{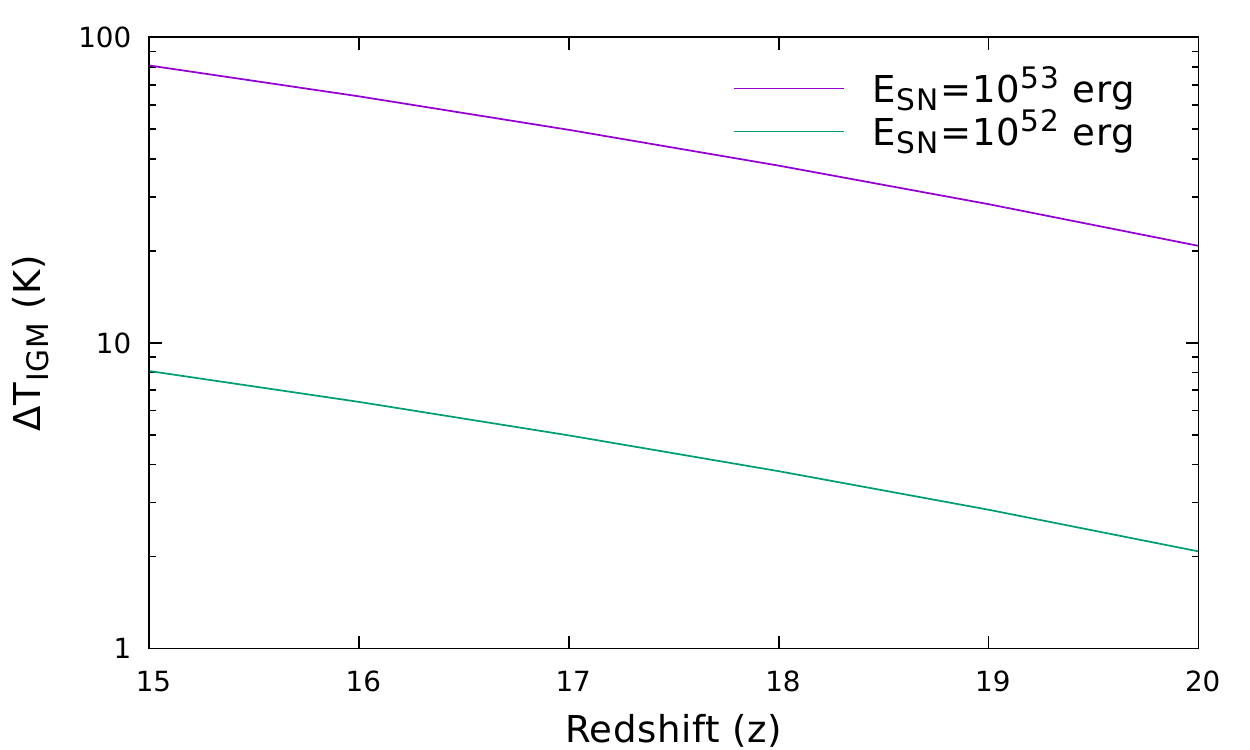}
\caption{Global increase in IGM temperature by CR protons for $E_{\rm SN}=10^{52}, 10^{53}$ erg as a function of redshift, without considering diffusion of CR. }
\label{fig:temp}
\end{figure}

This radio background is weakly dependent on the redshift of the occurrence of Pop III supernovae. If we use the corresponding comoving number density
of minihaloes at $z=20$ ($141 \, (h/0.6774)^3$ Mpc$^{-3}$), the redshift at which the observed absorption trough begins, then the emissivity is
$\approx 3.4 \times 10^{-43} \, {\rm erg} \, {\rm s}^{-1} \, {\rm Hz}^{-1} \, {\rm cm}^{-3} \, {\rm sr}^{-1} \zeta_{B,100}^{1.6} \, E_{53}$. Using the value of $c/H(z)$ at $z=20$
($83 \, (h/0.6774)$ Mpc),
one gets a brightness temperature,
\be
T_B (z=20) \approx 138  \, \zeta_{B,100}^{1.6} E_{53} \hbox{ }\rm K \,
\ee

The duration over which this radio background is produced will depend on the time distribution of Pop III supernovae, which is uncertain. In the simplest
scenario, if one considers these supernovae to go off within a short time (shorter than the corresponding Hubble time), then the radio background will amount
to the magnitude we have estimated above. At lower redshifts, this background will suffer cosmological dilution, and one can calculate the evolution
of the factor ${T_B+T_{\rm CMB}\over T_{\rm CMB}}$ with redshift, for different values of the initial redshift. We note that the constraints on cooling timescales as mentioned by \cite{sharma2018} is not relevant in our case because the radio background is not a sustained one.
\begin{figure*}
\centering
\includegraphics[height= 3in]{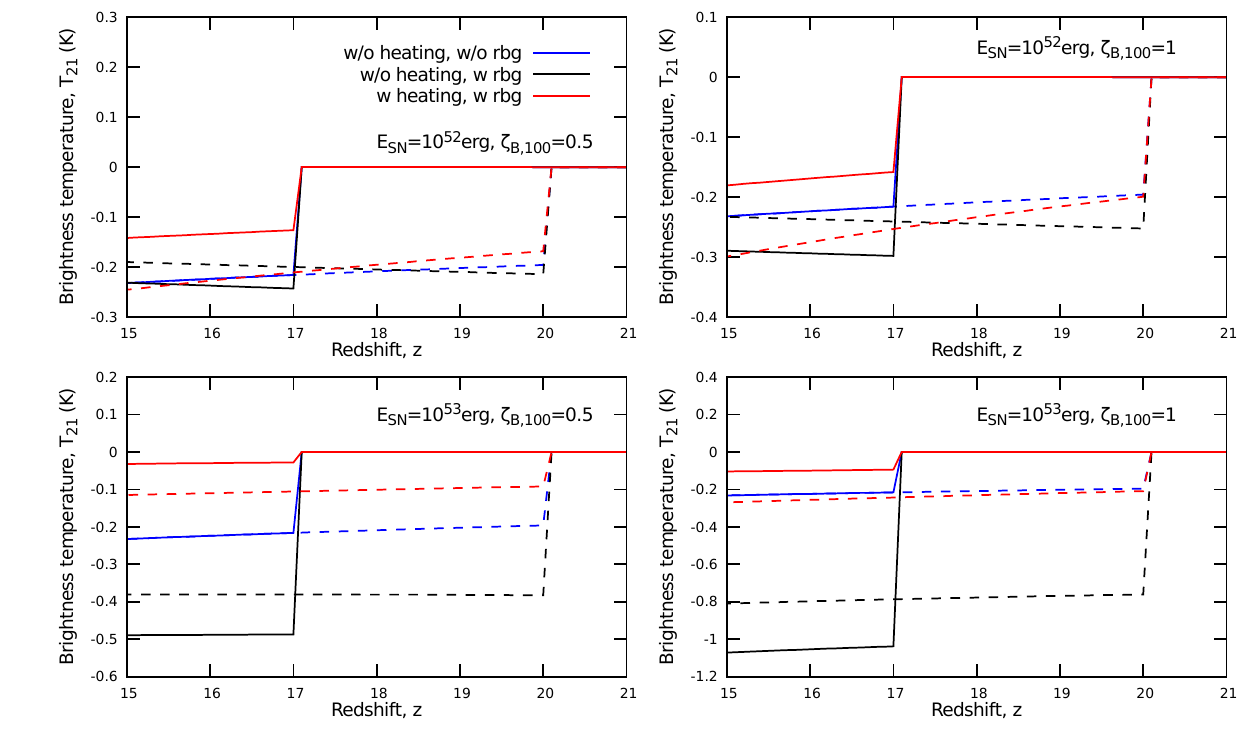}
\caption{Brightness temperature of 21 cm signal as a function of redshift, assuming a sudden coupling of spin temperature with gas temperature at $z=20$ (dashed lines), $17$ (solid lines), for three cases:
(1) one in the standard case, without any additional radio background and associated heating (blue lines), (2) with an additional radio background
but with heating suppressed (black lines), and (3) with a radio background and the associated heating  (red lines). The four panels show the cases with $E_{\rm SN}=10^{52}$ and $10^{53}$
erg, and $\zeta_{B,100}=1$ and $0.5$.}
\label{fig:optical-depth}
\end{figure*}

\section{Heating of IGM}
Another important implication of the present scenario is the heating of the IGM gas by CR protons, which would accompany the CR electrons
responsible for a radio background. \cite{sazonov2015} calculated the heating in this particular scenario, and we can use their result to discuss the
fraction of energy of Pop III supernovae that could go into CR protons. According to their estimate, protons with energy $\le 30$ MeV would lose
their energy within a Hubble time at $z\sim 20$. Suppose the fraction of energy contained in these low energy protons is $\eta_{LECR}$, then according
to their estimate, the temperature of the IGM gas would increase by (where $f_{\rm heat}=0.25$ is the fraction of CR proton kinetic energy deposited
as heat through Coulomb interactions),
\be
\Delta T_{\rm IGM}={2 f_{\rm heat} \eta_{LECR} \eta_{CR} E_{SN} \over  3 \, k \, (\rho_{\rm IGM} / 1.2 m_p)} n_h \,.
\ee
We show this temperature increase in Figure \ref{fig:temp} for $E_{\rm SN}=10^{53}$ erg,
$\eta_{CR}=0.15$, as used for the radio background calculation, and $\eta_{LECR}=0.05$ that is appropriate for $p=2.2$. The typical IGM gas temperature 
at $z\sim 20$ is $T \approx 2.725 \times 151 [ (1+z)/151]^{2}$ K$\approx 8$ K. The figure shows that the temperature increase due to CR protons can be substantial
and would go against deepening the absorption trough.  
However it is possible to obtain the same radio background as in equation 6 with $E_{53} <1$ 
as long as $\zeta_{B,100}$ is suitably larger than unity. This would also decrease the concomitant heating effect, as shown by the variation of heating
with $E_{\rm SN}$ in Figure \ref{fig:temp}.

In order to assess the important heating vis-a-vis the effect of the radio background due to Pop III SN explosions, we need to calculate the brightness temperature of the global 21 cm signal after the redshift of explosions. In the standard scenario, the spin temperature of neutral IGM gets decoupled from the CMB temperature at $z\sim 150$ as the gas decouples from CMB, and then gets coupled to it again at a lower redshift, due to lack of electrons. At a still lower temperature, due to the surge of Lyman-$\alpha$ photons coming from Pop III stars, the spin temperature gets coupled to the gas temperature again. For simplicity (and to describe our model in terms of the least number of free parameters), let us assume that the spin temperature $T_s$ becomes equal to $T_g$, the gas temperature at a certain redshift $z\sim 20$. Then the differential brightness temperature of 21 cm radiation is given by the standard expression (assuming the neutral fraction to be unity) \citep{zaldarriaga2004}.
\be
\Delta T_{21}(z) \approx 0.023 \hbox{ } K \Big[\Big(\frac{0.15}{\Omega_m h^2}\Big)\Big(\frac{1+z}{10}\Big)\Big]^\frac{1}{2} \Big(\frac{\Omega_b h^2}{0.02}\Big) \Big[1-\frac{T_R(z)}{T_S(z)}\Big] \,.
\ee
Here $T_R=T_B+T_{\rm CMB}$.
We show the resulting absorption feature in Figure \ref{fig:optical-depth}, without the effect of X-ray heating, but including the effect of heating due to protons from Pop III SNe.
We show three curves: (1) one in the standard case, without any additional radio background and associated heating, (2) with an additional radio background
but with heating suppressed, and (3) with a radio background and the associated heating, for various combination of parameters, for two cases in which the
coupling of spin temperature with gas temperature occurs at $z=20$ and $17$.

The curves in Figure \ref{fig:optical-depth} show that it is difficult to achieve the brightness temperature as observed by {\it EDGES} if the heating due to protons is taken into account. The heating is decreased when $E_{SN}$ is low, but then the radio background is also low, and the brightness temperature remains the same. Changing the cosmic ray efficiency parameter $\eta_{\rm CR}$ also does not make any difference. Decreasing $\zeta_{B,100}$ reduces the brightness temperature because of the decrease in the radio background intensity.  However, one could argue that the ratio of CR electrons to protons, which we have taken to be 
$(m_e/m_p)^{(3-p)/2}\approx 0.05$ could be different and it would change the radio background intensity without altering the heating effect. If this ratio is $\approx 0.25$ at $z=17$ (and $0.13$ at $z=20$) then one could get an absorption trough of depth $\approx 0.5$ K. The trade-off between these two effects is an important testable
prediction of the two-sided effect of Pop III SNe for future experiments.

\section{Discussions}
We can calculate the brightness temperature of the redshifted radio background at the present epoch, and compare with the observations by {\it ARCADE-2} experiment \citep{fixsen2011} (independently confirmed by \citet{Dowell2018}). 
The $1.4$ GHz signal when redshifted to present epoch, yields a brightness temperature of $160/18=8.9$ K at $78.89$ MHz, whereas the radio background
observed at the same frequency by {\it ARCADE-2} is $\approx 845$ K, two orders of magnitude higher than the background considered by us.
This conclusion is consistent with that of \citet{feng2018} who noted that a background with intensity even $1\%$ of the {\it ARCADE-2} result would have
an observable effect at high redshift $21$ cm observations. We note that one can increase the radio background intensity by increasing $f_{\rm SN}$, the number
of SN per minihalo, and bring it to agreement with the {\it ARCADE-2} (although $f_{\rm SN}\sim 100$ would be rather unrealistic), but then the IGM heating would scale up by a similar factor.

Another important prediction from the present model is the number of sources in the radio sky. 
The number of shells (before they are rapidly decelerated and stop
accelerating CRs) per solid angle is (where $r(z)=d_L(z)/(1+z)$ is the coordinate distance to redshift $z$ and $d_L$ is the luminosity distance),
\be
{c r(z)^2  \Delta z \times n_h \over H(z)} =c r(z)^2 n_h \, \Delta t \,(1+z) \approx 7 \times 10^{11} \, t_{14} \,.
\ee
Within the uncertainties of the estimate, this matches the limit of $\ge 6 \times 10^{11}$ sources per sr put by \citet{condon2012}.
We note here that  \citet{condon2012} derived the limit on source density using a beam size
of a few arc seconds, the limit can also be explained if the sources are extended.

The relativistic electrons that radiate at $1.4$ GHz at $z=17$ have a typical Lorentz factor of $4\times 10^3 \, \zeta_{B,100}^{-1/2}$. These electrons will
also inverse Compton scatter the CMB photons (with $T_{\rm CMB}=49$ K at $z=17$) to $\sim 66$ keV X-ray photons. The energy density of these X-ray
photons in the range $\sim 1\, - \,100$ keV is roughly $0.05$ eV cm$^{-3}$, which is a fraction $\sim 2 \times 10^{-6}$ of the CMB energy density at that redshift.
Therefore the diffuse X-ray background produced by the CR electrons (responsible for the proposed radio background) will not pose any problem with
observed X-ray background radiation (which is a fraction $\sim 10^{-3}$ of the CMB).

The corresponding $\gamma$-ray background produced by hadronic interactions of CRs with IGM protons is much below the observed background today.
The CR proton energy density is $\epsilon_{\rm CR} \approx 10^{-15}$ erg cm$^{-3}$ at $z=17$. Using a cross-section of $\sigma_{\rm pp}\sim 2.5 \times 10^{-26}$ cm$^2$, 
and IGM proton density $\sim 2.4 \times 10^{-3}$ cm$^{-3}$ at that redshift, one gets an energy density of $\gamma$-rays,
\be
{1\over 3}{ c \sigma_{\rm pp} \, n_{IGM} \, \epsilon_{\rm CR} \over   H(z)}\approx 6.4 \times 10^{-18} \, {\rm erg} \, {\rm cm}^{-3} \,.
\ee
When redshifted to present epoch, this amounts to an energy density of $\sim 6.1 \times 10^{-23}$ erg cm$^{-3}$, much below the observed diffuse $\gamma$-ray
background energy density of $\sim 10^{-18}$ erg cm$^{-3}$ \citep{ackermann2015}.

\section{Summary}
We have considered the effect of CRs accelerated during Pop III supernovae, which are an inevitable consequence of the emergence of the 
first stars in the Universe. Various effects of these CRs have been earlier considered in the literature, including the heating of IGM
and the neutrino background. We have shown that the radio background arising from CR electrons can be large enough to explain the recently
observed depth of absorption trough by {\it EDGES} but then the associated heating due to CR protons would tend to decrease the brightness temperature. Our model provides a testable prediction of this important process, and it is hoped that future experiments will shed more light on 
this era.

\section*{Acknowledgements}
We want to thank the referee, Dr Anastasia Fialkov, for the invaluable comments and suggestions which have helped to improve the manuscript.

\footnotesize{

\end{document}